\documentclass[usenatbib,usegraphicx]{mn2e}
\usepackage[T1]{fontenc}
\usepackage[latin9]{inputenc}
\usepackage{amsmath}
\usepackage{amssymb}
\usepackage{times}

\newcommand{\elCohMat}{\boldsymbol{\mathsf{\Gamma}}}
\newcommand{\briThree}{\boldsymbol{\mathsf{B}}^{(3)}}
\newcommand{\briTwo}{\boldsymbol{\mathsf{B}}}
\newcommand{\briTwoProj}[1]{\boldsymbol{\mathsf{B}}^{(#1)}}
\newcommand{\visThree}{\boldsymbol{\mathcal{V}}^{(3)}}
\newcommand{\visTwoProj}{\boldsymbol{\mathsf{R}}}
\newcommand{\vCZ}{\lowercase{v}\uppercase{C-Z}}

\title[Wide-field, polarimetric van Cittert-Zernike theorem]{A generalised
Measurement Equation and van Cittert-Zernike theorem for wide-field
radio astronomical interferometry}

\author[T. D. Carozzi and G. Woan]{
  T. D. Carozzi\thanks{E-mail: t.carozzi@physics.gla.ac.uk}
  and G. Woan\\
Dept. of Physics \& Astronomy, University of Glasgow, UK }

\begin{document}
\date{Received 2009 January 19; Revised 2008 October 31; in original form 2008 April 16}
\pagerange{\pageref{firstpage}--\pageref{lastpage}}\pubyear{2009}
\maketitle

\label{firstpage}

\begin{abstract}
We derive a generalised van Cittert-Zernike (vC-Z) theorem for radio 
astronomy that is valid for partially polarized sources over an
arbitrarily wide field-of-view (FoV).
The classical vC-Z theorem is the theoretical foundation of radio astronomical
interferometry, and its application is the basis of interferometric imaging.
Existing generalised vC-Z theorems in radio astronomy assume,  however,
either paraxiality (narrow FoV) or scalar (unpolarized) sources.
Our theorem uses neither of these assumptions, which are seldom fulfilled in
practice in radio astronomy, and treats the full electromagnetic field.
To handle wide, partially polarized fields, we extend the 
two-dimensional electric field (Jones vector) formalism of
the standard ``Measurement Equation'' of radio astronomical
interferometry to the full three-dimensional formalism developed in
optical coherence theory.
The resulting vC-Z theorem enables all-sky imaging in a single telescope pointing,
and imaging using not only standard dual-polarized interferometers (that measure 2-D electric fields),
but also electric tripoles and electromagnetic vector-sensor interferometers.
We show that the standard 2-D Measurement Equation is easily obtained from our formalism
in the case of dual-polarized antenna element interferometers.
We find, however, that such dual-polarized interferometers can have
polarimetric aberrations at the edges of the FoV that are often correctable.
Our theorem is particularly relevant to proposed and recently developed
wide FoV interferometers such as LOFAR and SKA, for which direction-dependent effects
will be important.
\end{abstract}
\begin{keywords}
telescopes; techniques: interferometric; techniques: polarimetric;
            instrumentation: interferometers;  instrumentation: polarimeters.
\end{keywords}

\section{Introduction}

Polarimetric wide-field imaging is a recent and important trend in
radio astronomy. Many new and planned radio telescopes such as LOFAR,
LWA, MWA, and SKA have a wide field-of-view (FoV) as a major feature.
Technologically, this has been made possible by the development of
phased dipole arrays and focal plane arrays, both of which have an
inherently wider FoV than traditional single-pixel radio telescopes.
The motivation for wide FoV polarimetric radio telescopes in astronomy
is that they facilitate the study of polarized phenomena
not restricted to narrow fields such as the large, highly-structured,
polarized features discovered in recent polarimetric galactic
surveys \citep{Taylor2006}.
With LOFAR now producing its first \emph{all-sky} images \citep{LOFARteam},
a new era of polarimetric wide-field imaging is starting in radio astronomy.

Despite this trend, a complete theory for wide-field, polarimetric astronomical
interferometry is lacking. Ultimately, classical interferometry is based
on the far-zone form of the van Cittert-Zernike (vC-Z) theorem
\citep{TMS01,Mandel95} which, in its original
form, is a scalar theory\footnote{By scalar theory we mean a theory
which only considers unpolarized radiation.}
and only valid for narrow fields. Such restrictions are obviously
unacceptable in astronomy where polarization often provides crucial astronomical
information, and sources are distributed on the celestial sphere, not necessarily
limited to small patches. 
In radio astronomy, a polarimetric, or vector (i.e. non-scalar), extension
of the narrow-field vC-Z theorem was given by \citet{Morris1964},
and more recently this was given a consistent mathematical
foundation by \citet{Hamaker2000} in the form of the
so-called \emph{Measurement Equation} (M.E.) of radio astronomy.
On the other hand, a wide-field extension of the scalar vC-Z theorem for
radio astronomy was given by \citet{Brouw1971}, 
and more recently, imaging techniques for scalar, wide-fields have been
an active field of
research \citep{Cornwell92,Sault1999b,Cornwell2005,McConnell06}.
However, a generalised vC-Z theorem for radio astronomy that is both
polarimetric and wide-field has not been derived.
This may be because it has been assumed that such a theorem
would be a trivial vector- or matrix-valued analogue of the wide-field,
scalar theory \citep{TMS01}.
As we will see, however, the final result is not that simple.

The purpose of the present work is therefore to derive a
vC-Z type relation by generalising the standard M.E. formalism to allow
for arbitrarily wide fields. This is achieved by generalising the two-component Jones
formalism to a three-component Wolf formalism \citep{Wolf54}.
In what follows, we will derive a vC-Z relation that is valid for the
entire celestial sphere and is fully polarimetric. We will show that
the standard M.E. can be recovered through a two-dimensional projection.
We will also show that a dual-polarized interferometer of short electric dipoles
(Hertzian dipoles) is inherently aberrated polarimetrically.
We also extend our vC-Z relation
to include not only the full electric field, but also the full magnetic
field, and thereby establish the complete second-order statistical
description of the electromagnetic relation between source terms and
interferometer response.

Our wide-field vC-Z relations should be of relevance to the subject
of \emph{direction-dependent effects} in radio interferometric imaging,
which has attracted recent attention \citep{Bhatnagar2008}.

\section{Deriving a \vCZ\ relation in three-component formalism}
\label{sec:WolfvCZ}
The vC-Z theorem as used in astronomy is different from its
original use in optical coherence. In astronomy one
images sources on the celestial sphere based on
localised measurements of their far-fields.
This is possible because the vC-Z theorem provides an explicit relationship
between the visibility measured directly
by an interferometer and the brightness distribution of the sources
\citep[chap 14]{TMS01}.
This makes the vC-Z relationship the foundation of synthesis mapping and
interferometric imaging.
Despite its importance, the vC-Z relations in use in astronomy are all
either based on the paraxial approximation (narrow FoV)
or they take the source emissions to be scalar.
These simplifications are questionable when the
sources cover wide fields or are highly polarized, as is often the case
in radio astronomy.

Here we derive the full vector electromagnetic analogue to the
wide-field, scalar vC-Z relation derived in \citet[chap 14]{TMS01},
and, as we will see, the final result is not simply a matrix- or vector-valued
version of the scalar relation. 
We seek a relationship between the electromagnetic field coherence of
a distribution of radio astronomical sources and the resulting electromagnetic
field coherence at a radio astronomical interferometer.
To simplify the discussion, we will use the term \emph{interferometer} as a shorthand
for radio astronomical interferometer\footnote{Most of the formalism also applies
to single-pixel radio telescopes as a special case when the interferometer
baseline length is zero.}. The treatment is intended for Earth-based
interferometer observations, but it also has space-based radio astronomical
interferometry in mind. 
\begin{figure*}
\begin{centering}
\includegraphics[clip,width=1\textwidth]{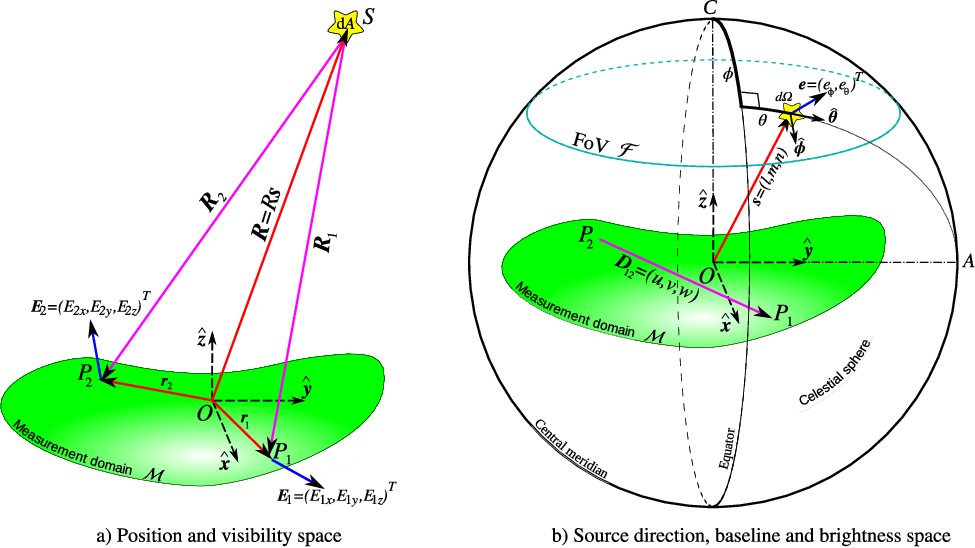}
\par
\end{centering}

\caption{\label{fig:CoordSys}Coordinate systems and notations used in the
text. a) A point source $S$ located in look-direction $\mathbf{s}$
at a distance $R$ with cross-section $\mathrm{d}A$ emits radiation, and its
electric field $\mathbf{E}$ is measured at the pair of
points $\mathbf{r}_1$ and $\mathbf{r}_2$ located within the
domain $\mathcal{M}$. At the approximate centre of $\mathcal{M}$ is
the origin of the coordinate systems $O$.
b) We use both a spherical and a Cartesian system to express the wide-field
vC-Z.
These systems are constructed with reference to $C$, the centre
of the FoV $\mathcal{F}$, and $A$ an arbitrary direction orthogonal to $C$
(which for Earth-based observations could be towards the North pole or zenith).
The right-handed Cartesian system $xyz$ is used for the source direction unit
vector $\mathbf{s}=l\mathbf{\hat{x}}+m\mathbf{\hat{y}}+n\mathbf{\hat{z}}$
and the relative displacement
vector $\mathbf{D}_{\lambda}=u\mathbf{\hat{x}}+v\mathbf{\hat{y}}
+w\mathbf{\hat{z}}$.
The basis vector $\mathbf{\hat{z}}$ points
towards $C$ and $\mathbf{\hat{y}}$ points toward $A$.
The angular part of the spherical coordinate system, $\{\phi,\theta\}$,
has its zero point at $C$, and the pole $A$ is at $\theta=+\pi/2$.
Relative to an arbitrary point on the celestial sphere
given by $\mathbf{s}$, the unit base vector $\boldsymbol{\hat{\theta}}$
is directed along the arc from the point towards $A$,
and $\boldsymbol{\hat{\phi}}$ points in the anticlockwise direction as seen
from $A$.
These base vectors are used to specify the source electric field
$\mathbf{e}=e_\phi\boldsymbol{\hat{\phi}}+e_\theta\boldsymbol{\hat{\theta}}$
and consequently also the two-dimensional brightness matrix.
}
\end{figure*}

Consider the problem illustrated in Fig. \ref{fig:CoordSys}a).
Radio emissions from far away sources, such as the point source $S$,
are measured by an interferometer located in the domain $\mathcal{M}$.
We want to establish a relationship between the coherence of the
electric field emanating from a source distribution and the
coherence of the electric field in $\mathcal{M}$.

Let us first consider a single point source $S$ located at $\mathbf{R}$.
The source is thus at a distance $R=|\mathbf{R}|$ in the direction given
by the unit vector (direction cosines) $\mathbf{s}=\mathbf{R}/R$  with
respect to the origin $O$ approximately at the centre of $\mathcal{M}$.
The source is within the interferometers FoV $\mathcal{F}$
which is centred on point $C$ on the celestial sphere.
The electric field from $S$ is measured by the interferometer
at pairs of positions $\mathbf{r}_1$ and $\mathbf{r}_2$ in $\mathcal{M}$.
$\mathcal{M}$ is assumed to be bounded, so the maximum distance between any two
measurement points $D_\mathrm{M}=\max_{\mathcal{M}}|\mathbf{r}_1-\mathbf{r}_2|$ (maximum baseline) is finite.
Although the source emission may be broadband we split
it into narrow, quasi-monochromatic, spectral bands and consider
a typical narrow (bandwidth much smaller than centre frequency) band centred on frequency $\nu$.
The assumption that $S$ is very far away, which
quantitatively we take to mean that $|\mathbf{R}-\mathbf{r}|\gg c/\nu$ for
all $\mathbf{r}\in \mathcal{M}$, implies that the entire interferometer is in
the far-field of the source.
This means that the electric field  $\mathbf{E}$ 
at point $\mathbf{r}_1$ at time $t$ is
\begin{equation}
  \mathbf{E}(\mathbf{r}_1,\mathbf{s},t)
   =\boldsymbol{\mathcal{E}} \left(\mathbf{\hat{R}}_1;\mathbf{s},
     t-\frac{R_1}{c}\right)
     \frac{e^{-\mathrm{i}2\pi\nu (t-R_1/c)} }{R_1}
    \label{eq:disElContribVec}
\end{equation}
where
\begin{equation}
R_1=|\mathbf{R}_1|=|\mathbf{r}_1-R\mathbf{s}|
 \label{eq:defRelPosSrcFld},\quad \mathbf{\hat{R}}_1=\mathbf{R}_1/R_1
\end{equation}
and $\boldsymbol{\mathcal{E}} \in \mathbb{C}^3$ is the complex
electric field amplitude vector emitted by the source at $\mathbf{s}$
in the direction of $\mathbf{r}_1$.

Furthermore, in the far-field, $\boldsymbol{\mathcal{E}}$ is approximately
transverse so that 
\begin{equation}
\mathbf{\hat{R}}_1\cdot\boldsymbol{\mathcal{E}} \approx 0 .
\end{equation}
If we further assume that the angular extent of $\mathcal{M}$ as
seen from $S$ is small, that is
\begin{equation}
|\mathbf{r}_1|\le D_M\ll R,
\end{equation}
then the interferometer is in the Fraunhofer far-field of the sources.
In this case we can use the approximations $\mathbf{\hat{R}}_1 \approx \mathbf{s} $
and $1/R_1\approx 1/R$ to simplify the expression for the electric field
in $\mathcal{M}$ to
\begin{equation}
  \mathbf{E}(\mathbf{r}_1,\mathbf{s},t)
   \approx \boldsymbol{\mathcal{E}} \left(\mathbf{s},
       t-\frac{R_1}{c}\right)
     \frac{e^{-\mathrm{i}2\pi\nu(t-R_1/c)} }{R}
    \label{eq:disElContribVecApprox}
 \end{equation}
where
\begin{equation}
  \mathbf{s}\cdot\boldsymbol{\mathcal{E}} \approx 0.
  \label{eq:transCondApprox}
\end{equation}
Here we have dropped the first argument of $\boldsymbol{\mathcal{E}}$ since
we are assuming that the angular variation of the sources emissions 
is small enough so that the $\mathbf{\hat{R}}_1 \approx \mathbf{s} $ approximation implies that 
$
  \boldsymbol{\mathcal{E}} \left(\mathbf{\hat{R}}_1;\mathbf{s},
      t\right)\approx
    \boldsymbol{\mathcal{E}} \left(\mathbf{s};\mathbf{s},t\right)
$. 

A similar expression to equation (\ref{eq:disElContribVecApprox})
for the field at $\mathbf{r}_2$ is obtained by replacing $\mathbf{r}_1$ 
with $\mathbf{r}_2$  and $R_1$ with $R_2$.
The electric coherence matrix (tensor) $ \elCohMat $ can then be found by taking the
outer product of the electric fields
\begin{equation}
\mathsf{\Gamma}_{ij}=
  \left\langle  E_i(\mathbf{r}_1,t) E_j^\ast(\mathbf{r}_2,t)\right\rangle
  \label{eq:EcohMat}
\end{equation}
where $E^\ast$ denotes complex conjugation,
$\langle\;\rangle$ denotes time averaging,
and the subscripts $i,j$
label Cartesian components $x,y,z$, which we will define more precisely later.
We have written the electric coherence matrix, equation (\ref{eq:EcohMat}),
as a $3\times 3$ complex matrix even though, for the single point source we are
considering here, its
rank is two and could therefore be expressed as a $2\times 2$ complex matrix.
We keep the electric coherence matrix as a $3\times 3$ complex
matrix since it is valid even when there are more than one point
source.

We now move to the case of a finite number of point sources.
The total electric field measured at points  $\mathbf{r}_1$ and  $\mathbf{r}_2$
is now the sum of the fields from sources in directions $\mathbf{s}_p$
for $p=1,2,...$. The electric coherence matrix is therefore
\begin{align}
 \left\langle  E_i(\mathbf{r}_1,t)
   E_j^\ast(\mathbf{r}_2,t)\right\rangle &
  =  \sum_{p}\sum_{q} \left\langle  E_i(\mathbf{r}_1,\mathbf{s}_p,t)
   E_j^\ast(\mathbf{r}_2,\mathbf{s}_q,t)\right\rangle \nonumber \\
     &\approx  \sum_p \left\langle \mathcal{E}_i 
       \left(\mathbf{s}_p,t\right)
     \mathcal{E}_j^\ast \left(\mathbf{s}_p,t\right)
          \right\rangle
     \frac{e^{-\mathrm{i}2\pi\nu\mathbf{s}\cdot
      (\mathbf{r}_1-\mathbf{r}_2)/c} }{R^2}
  \label{eq:discreteSrcEcorr}
\end{align}
for $i,j=x,y,z$.
In going from the double to the single sum we used the usual vC-Z assumption
that the sources are spatially incoherent,
that is, sources in different directions are statistically independent.

Until now we have considered only discrete sources. We can make the transition
to the more general continuum source distribution by introducing the
three-dimensional brightness matrix $\briThree $ as a function of
direction $\mathbf{s}$ in a continuous source distribution
\begin{equation}
\sum_p\left\langle \mathcal{E}_i \left(\mathbf{s}_p,t\right)
     \mathcal{E}^\ast_j \left(\mathbf{s}_p,t\right)
          \right\rangle
   \rightarrow \int_\mathrm{source}\mathsf{B}^{(3)}_{ij}(\mathbf{s}) \,\mathrm{d}A ,
   \label{eq:defBriThree}
\end{equation}
where  $\mathrm{d}A$ is the infinitesimal area of the source distribution.
The superscript is to highlight the fact that  $\briThree $ is three-dimensional
as opposed the usual two-dimensional brightness matrix \citep{Hamaker2000}.
The reason that the brightness matrix here is three-dimensional is simply because the
full electric field amplitude $\boldsymbol{\mathcal{E}}$ is three-dimensional.
However, due to equation (\ref{eq:transCondApprox}), not all the components of  $\briThree $
are arbitrary. In fact, we will show that it can be recast as one
two-dimensional matrix.

By making the replacement (\ref{eq:defBriThree}) in
equation (\ref{eq:discreteSrcEcorr}), we obtain
\begin{align}
  \elCohMat(\mathbf{r}_1,\mathbf{r}_2,0) 
    & =\int_{\mathrm{source}}\briThree(\mathbf{s})
      \frac{\mathrm{e}^{-\mathrm{i}2\pi\nu\mathbf{s}\cdot
          (\mathbf{r}_1-\mathbf{r}_2)/c }
      }{R^2}\,\mathrm{d}A \nonumber\\
          & =\int_{\mathcal{F}}\briThree(\mathbf{s})
      \mathrm{e}^{-\mathrm{i}2\pi\nu\mathbf{s}\cdot
          (\mathbf{r}_1-\mathbf{r}_2)/c }
      \,\mathrm{d}\Omega ,
    \label{eq:continumize}
\end{align}
where $\mathrm{d}\Omega$ is an infinitesimal solid angle of the source
distribution in $\mathcal{F}$.
In this expression we see that the dependence on the pairs of position is
only relative, that is, the 
coherence matrix depends only on $\mathbf{r}_1-\mathbf{r}_2=\mathbf{D}$,
and so we recast the expression in terms of the vector
\[
  \mathbf{D}_{\lambda}=\frac{\nu}{c}(\mathbf{r}_1-\mathbf{r}_2) ,
\]
also known as the baseline vector measured in
wavelengths ($\lambda=c/\nu$). 

In practice, rather than use $ \elCohMat$ directly, it is convenient and conventional
to put the phase reference point at the centre of the FoV given by the direction $\mathbf{s}_0$.
The result of the change in phase,
\[
   \visThree(\mathbf{D}_{\lambda})=\exp(\mathrm{i}2\pi \mathbf{s}_0
    \cdot\mathbf{D}_{\lambda})
    \elCohMat_{\nu}(\mathbf{r}_{1},\mathbf{r}_{2}),
\]
is the 3$\times$3 generalisation of the
standard 2$\times$2 visibility matrix as defined by,
for instance, \citet{Hamaker2000}.  $\visThree$ is a general complex
matrix except for $\mathbf{D}_\lambda=0$ where it is Hermitian.
It fulfill the symmetry relation
$\visThree(-\mathbf{D}_\lambda)=\left(\visThree(\mathbf{D}_\lambda)\right)^\dagger$
where $^\dagger$ stands for Hermitian transpose.

If we use $\visThree$ in equation (\ref{eq:continumize}) we arrive at 
\begin{equation}
  \visThree(\mathbf{D}_{\lambda})=\int_{\mathcal{F}}\briThree(\mathbf{s})
     \mathrm{e}^{-\mathrm{i}2\pi(\mathbf{s}-\mathbf{s}_0)
       \cdot\mathbf{D}_{\lambda} }
      \,\mathrm{d}\Omega .
  \label{eq:vCZ3Dimpl}
\end{equation}

Equation (\ref{eq:vCZ3Dimpl}) is a matrix version of the wide-field,
scalar vC-Z relation and so includes the partial polarization of
the (non-scalar) source distribution. However, it is not very useful 
in this form since it does not automatically fulfill the constraint
(\ref{eq:transCondApprox}). When applied to $\briThree$, this
constraint becomes
\begin{equation}
   \briThree \mathbf{s}=\mathbf{0}
  \label{eq:B3constr}
\end{equation}
for all directions $\mathbf{s}$, where $\mathbf{s}$ is understood to be a column vector.  
The constraint can, however, easily be removed if we
express $\boldsymbol{\mathcal{E}}$
in terms of spherical base vectors and set the radial component
to zero. This leads us to introduce two coordinate systems: a Cartesian
and a spherical.

We use the angular, or tangential, basis
set $\{\boldsymbol{\hat{\phi}},\boldsymbol{\hat{\theta}}\}$ of a spherical
polar coordinate system as the basis for the polarization of the transverse
field of the source distribution on the celestial sphere, see Fig. 1b).
To simplify the results, the spherical coordinate system is taken relative to the
phase reference position of the interferometer assumed to be at $C$.
In other words,
the zero point of the spherical system, $\{\phi,\theta\}=\{0,0\}$,
(intersection of the equator and central meridian),
is taken to coincide with $C$.
$\theta $ is the angle from the equator (positive in the hemisphere with the
pole $A$ and negative in the other hemisphere), and $\phi$ is the position angle
from the central meridian around $AO$ in the anticlockwise sense looking along $AO$.
In lieu of any other reference directions, the orientation of the spherical
system around $C$ is arbitrary, but
for Earth-based measurements $A$ could be directed towards the North
pole or zenith.

One should note that $\{\boldsymbol{\hat{\phi}},\boldsymbol{\hat{\theta}}\}$
are consistent with Ludwig's second definition
as detailed in \citet{Ludwig73}
with the understanding that the antenna boresight in \citet{Ludwig73} is here
at $C$, that
Ludwig's reference polarization unit vector $\hat{i}_{\mathrm{ref}}$ is here $\boldsymbol{\hat{\theta}}$, and
Ludwig's cross polarization unit vector $\hat{i}_{\mathrm{cross}}$ is here $\boldsymbol{\hat{\phi}}$.
See \citep{Piepmeier04} for the use of Ludwig's third definition in a vC-Z relation.

The Cartesian system, with base
vectors $\{\mathbf{\hat{x}},\mathbf{\hat{y}},\mathbf{\hat{z}}\}$, is defined
so that $\mathbf{\hat{z}}$ is in the direction of $C$, and $\mathbf{\hat{y}}$
is in the direction of the pole $A$.
In terms of the Cartesian system, we can explicitly write
the components of the vectors in equation (\ref{eq:vCZ3Dimpl}) as
\begin{align}
  \mathbf{s}    &=l\mathbf{\hat{x}}+m\mathbf{\hat{y}}+n\mathbf{\hat{z}}=(l,m,n)^{\mathrm{T}},\label{eq:s_lmn}\\ 
  \mathbf{s}_{0}&=\mathbf{\hat{z}}=(0,0,1)^{\mathrm{T}},\label{eq:s0_lmn}\\ 
  \mathbf{D}_{\lambda}     &=u\mathbf{\hat{x}}+v\mathbf{\hat{y}}+w\mathbf{\hat{z}}=(u,v,w)^{\mathrm{T}} \label{eq:D12_uvw}
\end{align}
where
\begin{equation}
   n=\pm\sqrt{1-l^2-m^2}
\end{equation}
and where the superscript $^\mathrm{T}$ stands for vector transpose.
All these vectors are unit vectors with real-valued components
ranging between $-1$ and $+1$.
These definitions of the $uvw$ and $lmn$ spaces are the same as the usual
definitions for Earth-based observations, see e.g., \citet{TMS01}.
Also the matrices in equation (\ref{eq:vCZ3Dimpl}) are to be considered in what
follows as being expressed in the Cartesian system. 

The relationships between the spherical
and Cartesian systems base vectors are
\begin{align}
  \boldsymbol{\hat{\phi}} &=  \frac{1}{\sqrt{1-m^2}}\left(n \mathbf{\hat{x}} 
     -l\mathbf{\hat{z}}\right) \label{eq:phixyz}\\
 \boldsymbol{\hat{\theta}} &= \frac{1}{\sqrt{1-m^2}} \left( -lm \mathbf{\hat{x}}+
   (1-m^2)\mathbf{\hat{y}}
    -mn  \mathbf{\hat{z}}\right) .  \label{eq:thetaxyz}
\end{align}

Using these spherical and Cartesian systems we can express
the three-dimensional transverse electric field as
\begin{equation}
  \boldsymbol{\mathcal{E}} =\boldsymbol{\mathsf{T}}
   \mathbf{e}
  \label{eq:CartAsTrans}
\end{equation}
where
\[
  \mathbf{e}(l,m)=\left(\begin{array}{c}
  e_{\phi}(l,m)\\
  e_{\theta}(l,m)
  \end{array}\right)
\]
is the Jones vector in spherical (rather than the usual Cartesian) components and
\begin{equation}
\boldsymbol{\mathsf{T}}=\frac{1}{\sqrt{1-m^{2}}}\left(\begin{array}{cc}
n   & -lm  \\
0   & 1-m^2\\
-l & -mn   \end{array}\right)\quad\textrm{for }m^{2}\neq1
  \label{eq:sph2cart_lmn}
\end{equation}
is the $3\times 2$ transformation matrix between the components given by
the equations (\ref{eq:phixyz}) and (\ref{eq:thetaxyz}).
Note that this transformation is possible for all directions on the celestial
sphere except for $m=\pm1$, i.e., the poles of the spherical coordinate
system.

If one wishes to use to a polar spherical system in which the centre of the FoV is not on the
equator, as it is here, but rather at some declination $\Theta$, then one simply replaces $\boldsymbol{\mathsf{T}}$
with $\boldsymbol{\mathsf{T}}'$ defined in appendix \ref{sec:CoordTrans}, equation (\ref{eq:sph2cart_gen}).
This assumes that the pole $A$ is towards the Earth's North pole.

It is easy to show that
\[
  \mathbf{s}^\mathrm{T}\boldsymbol{\mathsf{T}}\mathbf{e}=0 ,
\]
so the transverse electric field expressed according to
equation (\ref{eq:CartAsTrans}) does indeed
fulfill equation (\ref{eq:transCondApprox}). So if we use $\boldsymbol{\mathsf{T}} \mathbf{e}$
rather than $\boldsymbol{\mathcal{E}}$
in equation (\ref{eq:vCZ3Dimpl}) we would have an unconstrained vC-Z equation.
We can introduce this replacement by rewriting $\briThree$ using equation (\ref{eq:CartAsTrans}), so
\begin{equation}
  \briThree=\boldsymbol{\mathsf{T}}\briTwo
    \boldsymbol{\mathsf{T}}^{\mathrm{T}}
 \label{eq:Bri3FromBri2}
\end{equation}
where,  suppressing the dependence on $(l,m)$,
\begin{align}
\briTwo= & \left(\begin{array}{cc}
\left\langle \left|e_{\phi}\right|^{2}\right\rangle  
          & \left\langle e_{\phi}e_{\theta}^\ast\right\rangle \\
\left\langle e_{\theta}e_{\phi}^\ast\right\rangle 
    & \left\langle \left|e_{\theta}\right|^{2}\right\rangle \end{array}\right)
\end{align}
is the 2$\times$2 brightness matrix, but in spherical rather than Cartesian
coordinates.
By this we mean that, for an arbitrary direction $(l,m)$,
$\briTwo$ is locally equivalent to
the usual paraxial brightness matrix in Cartesian coordinates.
From its definition it easy to see that $\briTwo$ is a Hermitian matrix.

By using the $lmn$ and $uvw$ spaces, as spanned by the vectors $\mathbf{s}$,
and $\mathbf{D}_\lambda$, we can write equation (\ref{eq:vCZ3Dimpl})
in a more explicit form. The exact form, though, depends on the extent of $\mathcal{F}$.
If it is entirely in the hemisphere $n>0$, then we write  $\mathcal{F}=\mathcal{F}_+$ and
\begin{equation}
  \visThree(u,v,w)=\mathop{\iint}_{\mathcal{F}_+}\boldsymbol{\mathsf{T}}
              \briTwo\boldsymbol{\mathsf{T}}^{\mathrm{T}}
    \frac{\mathrm{e}^{-\mathrm{i}2\pi\left[ul+vm+w\left(n
     -1\right) \right]}}{n} \,\mathrm{d}l\mathrm{d}m ,
  \label{eq:vCZ3D}
\end{equation}
where we have used $\mathrm{d}\Omega=\mathrm{d}l\mathrm{d}m/|n|$,
and where the matrices depend implicitly on $l$ and $m$.

If, however, part of $\mathcal{F}$ is in the $n<0$ hemisphere, then we must add
to equation (\ref{eq:vCZ3D}) the contribution from this hemisphere given by the integral
\begin{equation}
  \mathop{\iint}_{\mathcal{F}_-}\boldsymbol{\mathsf{T}}
              \briTwo(l,m;n<0)\boldsymbol{\mathsf{T}}^{\mathrm{T}}
    \frac{\mathrm{e}^{-\mathrm{i}2\pi\left[ul+vm+w\left(n
     -1\right) \right]}}{|n|} \,\mathrm{d}l\mathrm{d}m ,
  \label{eq:vCZ3Dmin}
\end{equation}
where $\mathcal{F}_-$ is the subset of $\mathcal{F}$ in the $n<0$ hemisphere.

The image horizon, $n=0$, can also be included by
reparametrising the integral in terms of $(m,n)$ rather than $(l,m)$ and
using the replacement $\mathrm{d}\Omega=\mathrm{d}m\mathrm{d}n/|l|$.
The poles $m=\pm 1$ can also be imaged, but one must then stipulate the orientation
of the $\boldsymbol{\hat{\phi}}$ and $\boldsymbol{\hat{\theta}}$ vectors at
these singular points.

Now, by extending $\mathcal{F}_+$ to cover the entire $n>0$ hemisphere
and extending $\mathcal{F}_+$ to cover the entire  $n<0$ hemisphere, 
the entire celestial sphere can be imaged in a single telescope pointing.
An assumption here is that $\mathcal{M}$ is a proper three-dimensional
volume, or in other words, the baselines should be non-coplanar.
If $\mathcal{M}$ is just a plane (coplanar baselines) then only one hemisphere
can be mapped uniquely.

Equation (\ref{eq:vCZ3D}) is our main result.
It says that the full $3\times 3$ electric visibility
matrix $\visThree $ on the three-dimensional $uvw$ space
is given by the $2\times 2$ brightness matrix $\briTwo $ on the
two-dimensional $lm$ plane. The fact that this is a relationship between
two matrices with different matrix dimensions is a fundamental feature of our vC-Z,
and makes it clear that it is not just a matrix-valued generalisation of the
wide-field, scalar vC-Z. Mathematically, this is ultimately due to the ranks of the
fundamental matrices, of which we will speak more in section \ref{sec:stokesParams}. 

The remaining vC-Z relationships that form a complete characterisation of the
electromagnetic coherence response of a radio astronomical interferometer are given
in appendix \ref{EMvCZ}.

\section{Wide-field \vCZ\ relation as a transform}

In the previous section we derived a vC-Z relation, equation (\ref{eq:vCZ3D}),
in which the visibility matrix is determined from the brightness matrix.
It is well known that the original vC-Z theorem (far-zone form) for narrow-fields
states that there is a two-dimensional Fourier transform relationship between visibility
and brightness. In astronomical interferometry the transform aspect of this relationship
is exploited to produce brightness images from measured visibility.
The wide-field vC-Z, equation (\ref{eq:vCZ3D}), is not a two-dimensional Fourier transform,
but, as we will now show, it is still possible invert it and thereby establish a sort of
generalised transform.

First we should state that there are several ways of
expressing $\briTwo $ in terms of $\visThree $, even though the wide-field vC-Z relation,
equation (\ref{eq:vCZ3D}), is a one-to-one relationship in general.
This is because $\visThree $
has redundancies, as one can expect considering the 
asymmetry in the respective matrix dimensions of
$\briTwo $ and $\visThree $. So although the
rank of $\visThree$ is three in general, it is overdetermined if $\briTwo $
is known. 
We will first derive a
solution that is valid for the entire celestial sphere based on
the full $\visThree$.
The case when only a projection of  $\visThree$ is
available will be discussed in section \ref{sec:Dual-Pol-Int}.

Consider that we are given the full $\visThree (u,v,w)$ and that we would
like to solve equation (\ref{eq:vCZ3D}) for $\briTwo $.
By extending the solution of
the scalar problem described in \citet{Cornwell92} to the three-dimensional
matrix relationship in equation (\ref{eq:vCZ3D}) we find that one approximate
solution is
\begin{align}
  &\briThree(l,m;n>0)=  \sqrt{1-l^2-m^2}\;\times\nonumber \\
  &\;\int_0^1\mathop{\iiint}_{\mathcal{M}-\mathcal{M}'} \visThree(u,v,w)
   e^{\mathrm{i}2\pi[ul+vm+w(n-1)]}
   \, \mathrm{d}u\mathrm{d}v\mathrm{d}w \mathrm{d}n.
  \label{eq:vCZ3Dinv}
\end{align}
where $\mathcal{M}-\mathcal{M}'$ is the $uvw$ space spanned by $(\mathbf{r}-\mathbf{r}')/\lambda$
for $\mathbf{r}\in \mathcal{M}$ and $\mathbf{r}' \in \mathcal{M}$.
From $\briThree$, we can find a solution for the two-dimensional brightness matrix,
\begin{equation}
  \briTwo (l,m; n>0)=\boldsymbol{\mathsf{T}}^{\mathrm{T}}\briThree\boldsymbol{\mathsf{T}}.
  \label{eq:Bri2FromBri3}
\end{equation}
An analogous expression applies for $\briThree (l,m;n<0)$ but
with the integration over $n$ running from $-1$ to $0$ rather
than  $0$ to $1$. For $n=0$, an expression can be obtained by using
the $(m,n)$ parametrised vC-Z mentioned in the previous section.

We can now state the generalised transform as
\begin{equation}
  \visThree (u,v,w) \leftrightarrows\briTwo (l,m)
  \label{eq:vCZ3Drel}
\end{equation}
where $\leftrightarrows $ reads ``is wide-field, polarimetric vC-Z related to''.
The relation from brightness
matrix to visibility matrix is given (for $n>0$) by
equation (\ref{eq:vCZ3D}), and the relation from visibility matrix to
brightness matrix is given by equations (\ref{eq:Bri2FromBri3})
and (\ref{eq:vCZ3Dinv}). Note that the $\leftrightarrows $ relation
is not simply a two-dimensional Fourier transform as in the narrow-field
case. Furthermore, the difference in the dimensionality of $\visThree $
and $\briTwo $ make it clear that equation (\ref{eq:vCZ3Drel}) cannot simply
be a matrix generalisation of the scalar vC-Z relation, as is sometimes assumed.

\section{Using Stokes parameters for wide fields}
\label{sec:stokesParams}

In practice it is common to use Stokes parameters to characterise the brightness
and visibility matrices over narrow fields.
Let us see how Stokes parameters
can be applied to the wide-field vC-Z relation, equation (\ref{eq:vCZ3D}).

Let us first consider brightnesses. The difference between
the two-dimensional brightness matrix  $\briTwo$ in equation (\ref{eq:vCZ3D})
and the usual two-dimensional brightness matrix in the paraxial approximation
is that the former is defined on a spherical domain while the latter is defined
on a Cartesian plane. Locally, for a source at some $(l,m)$, the two brightness
matrices can be made equal. Thus we can
define the Stokes parameters in terms of the components of $\briTwo$ with respect
to a spherical basis  $\{\boldsymbol{\hat{\phi}},\boldsymbol{\hat{\theta}}\}$ in
analogy with Cartesian basis as 
\begin{equation}
 \mathbf{S}=\left(\begin{array}{c}
 I\\
 Q\\
 U\\
 V\end{array}\right)=\left(\begin{array}{c}
 B_{\theta\theta}+B_{\phi\phi}\\
 B_{\theta\theta}-B_{\phi\phi}\\
 B_{\theta\phi}+B_{\phi\theta}\\
 +\mathrm{i}\left(B_{\theta\phi}-B_{\phi\theta}\right)\end{array}\right).
 \label{eq:StokesBright}
\end{equation}
When the spherical system is aligned as it is in section \ref{sec:WolfvCZ}
such that the intersection of its central meridian and its equator is located
at the centre of the FoV and the pole $A$ is towards the Earth's North pole,
then in a sufficiently narrow field around the centre of the FoV
the Stokes brightnesses $(I,Q,U,V)^\mathrm{T}$ are approximately
equal to the Stokes parameters of the IAU \citep{Hamaker2}.
If one wishes to conform with IAU Stokes parameters over the entire celestial sphere,
then one can use equation (\ref{eq:StokesBright}) with $\boldsymbol{\mathsf{T}}$ replaced
by $\boldsymbol{\mathsf{T}}'(\Theta)$, equation (\ref{eq:sph2cart_gen}), where $\Theta $
is the declination of the centre of the FoV.
The correspondence is then that $+\boldsymbol{\hat{\theta}}$ is the IAU's $+\mathbf{\hat{x}}$,
and $+\boldsymbol{\hat{\phi}}$ is the IAU's $+\mathbf{\hat{y}}$.

Note however that the IAU basis set for polarimetry
is not the same as the Cartesian basis set
widely used in radio interferometry
for defining source directions and baselines, which in this paper is denoted
$\{\mathbf{\hat{x}},\mathbf{\hat{y}},\mathbf{\hat{z}}\}$.
In adopting both these systems, therefore, a sacrifice must be made,
and we have chosen to slightly modify the usual
Pauli matrices that are used to relate the brightness matrix to the Stokes
parameters. Explicitly, the brightness matrix in terms of the Stokes
parameters in equation (\ref{eq:StokesBright}) is in this paper
\begin{equation}
  \briTwo = \frac{1}{2}\left( \begin{array}{cc} I-Q & U+\mathrm{i}V \\
                 U-\mathrm{i}V &     I+Q \end{array} \right) .
  \label{eq:briTwoInStokes}
\end{equation}
As can be seen, the expansion of this expression into unitary matrices,
as in \citet[eq. 6]{Hamaker2000}, leads to matrices equivalent to Pauli
matrices but with a change of overall sign for the matrices associated
with $Q$ and $V$. 

In terms of equation (\ref{eq:StokesBright}), we can write
the three-dimensional brightness 
matrix in equation (\ref{eq:vCZ3D}) as
\begin{align}
 \briThree = \boldsymbol{\mathsf{T}}\briTwo\boldsymbol{\mathsf{T}}^{\mathrm{T}}
  = & I\boldsymbol{\mathsf{C}}_{I}+Q\boldsymbol{\mathsf{C}}_{Q}
     +U\boldsymbol{\mathsf{C}}_{U}+V\boldsymbol{\mathsf{C}}_{V} ,
  \label{eq:B3inStokes}
\end{align}
where
\begin{align*}
  \boldsymbol{\mathsf{C}}_{I}= & \frac{1}{2}\left(\begin{array}{ccc}
  1-l^{2} & -lm & -ln\\
  -lm & 1-m^{2} & -mn\\
  -ln & -mn & l^{2}+m^{2}\end{array}\right)\\
  \boldsymbol{\mathsf{C}}_{Q}= & \frac{1}{2}\left(\begin{array}{ccc}
  \frac{-n^{2}+l^{2}m^{2}}{1-m^{2}} & -lm & \frac{(1+m^{2})ln}{1-m^{2}}\\
  -lm &  1-m^{2} & -mn\\
  \frac{(1+m^{2})ln}{1-m^{2}} & -mn & \frac{-l^{2}+m^{2}n^{2}}{1-m^{2}}\end{array}
    \right)\\
  \boldsymbol{\mathsf{C}}_{U}= & \frac{1}{2}\left(\begin{array}{ccc}
  -2\frac{lmn}{1-m^{2}} & n & \frac{m(l^{2}-n^{2})}{1-m^{2}}\\
  n & 0 &  -l \\
  \frac{m(l^{2}-n^{2})}{1-m^{2}} & -l & 2\frac{lmn}{1-m^{2}}\end{array}\right)\\
  \boldsymbol{\mathsf{C}}_{V}= & \mathrm{i}\frac{1}{2}\left(\begin{array}{ccc}
   0 &  n & -m\\
  -n &  0 &  l\\
   m & -l &  0\end{array}\right).
\end{align*}
It is easy to see that the  $\boldsymbol{\mathsf{C}}_i$ for $i=I,Q,U,V$
depend only on $(l,m,n)$, and that they become the three-dimensional analogs
of the equivalent Pauli matrices in equation (\ref{eq:briTwoInStokes}).

Equation (\ref{eq:B3inStokes}) shows that for every look-direction $(l,m)$,
the three-dimensional brightness matrix has
four degrees of freedom, here expressed as the four Stokes parameters.
In other words, the four Stokes parameters completely characterise the partially
polarized brightness also for wide fields, under the assumptions made in the derivation
of equation (\ref{eq:vCZ3D})).

Now let consider if Stokes visibilities can be extended to wide fields.
A consequence of the three-dimensionality of $\visThree $ in equation (\ref{eq:vCZ3D}) is, 
however, that the Stokes parameters cannot in general fully
characterise the electric visibility.
This is because, in contrast to $\briThree $, which has a rank of at most two due to the
transversality condition\footnote{That equation (\ref{eq:B3constr}) implies that $\briThree$
has rank two comes from the rank-nullity theorem of linear algebra, since
equation (\ref{eq:B3constr}) implies that dimension of the null space is one and the matrix
dimension of $\briThree $ is three, so $3-1=2$ is the rank of  $\briThree $. },
equation (\ref{eq:B3constr}),  $\visThree $ has no similar constraint. 
Indeed one can convince oneself of the full rank of $\visThree $
by considering two distinct point sources: the weighted
sum of their $\briThree $ matrices at some $(u,v,w)$ point
according to equation (\ref{eq:vCZ3D}), will in general be rank three.
As there are only four Stokes parameters, albeit complex-valued in the case of visibilities,
they cannot fully parametrise a rank three matrix.

Alternatively, rather than use the standard four Stokes parameters,
one could also use complexified versions of the nine, real, generalised Stokes
parameters \citep{Carozzi00a}.
These parameters are analogous to the standard Stokes parameters but can
completely describe the coherence of the full three-dimensional electric field.
In light of the discussion above, these generalised Stokes parameters are
particularly suitable for parametrising $\visThree $, but we will not discuss them here any
further.

\section{Generalised Measurement Equations}
\label{sec:Dual-Pol-Int}

The vC-Z theorem is a basic, fundamental physical relationship
is independent of technology.
The measurement equation (M.E.) of radio astronomy,
on the other hand, includes practical aspects of
telescope measurements,
in particular the instrumental response of the telescope.
Usually it is a relationship between a 2$\times$2 brightness matrix
and a 2$\times$2 cross-correlation matrix
of the output-voltage of a dual-polarized interferometer.
Since in the past such two-dimensional M.E. have been tacitly based on the
paraxial approximation valid only for narrow fields, it is important
to verify that it can be recovered from the wide-field, polarimetric vC-Z,
equation (\ref{eq:vCZ3D}), for which the paraxial approximation is not used.
Although it is possible to do this in a simple, straightforward way,
we choose to do it in a more detailed way, introducing a
formalism that extends the usual two-dimensional, electric field based model
of radio astronomical antenna response.

\subsection{Electromagnetic antenna response model}
To obtain a 2-D M.E. we must first introduce a formalism for converting
the full electric field to a voltage in the interferometer antenna.
An electric field $\mathbf{E}$ at an antenna excites
an open circuit voltage $V$. Assuming linearity, these two quantities are
related as
\begin{equation}
  \label{eq:antEffLen}
  V=\mathbf{L}\cdot\mathbf{E}
\end{equation}
where $\mathbf{L}$ is the antenna effective length vector.   In general
it is a function of incidence direction, i.e. $(l,m)$, but here we will only
use ideal, Hertzian dipole antennas (short electric dipoles).
These have the important property
that their effective length does not depend on incidence direction, $\mathbf{L}$ is just
a constant unit vector, and so it directly samples the component of the
electric field along its length.

If we have $n$ co-located antennas that have no mutual coupling, their output
voltages can be written in a matrix form
\begin{equation}
  \label{eq:antEffLenSet}
  \left(\begin{array}{c} V_1 \\ V_2 \\ \vdots \\ V_n \end{array}
  \right)
   =\left(\begin{array}{ccc}L_x^{(1)} &L_y^{(1)} &L_z^{(1)} \\
                            L_x^{(2)} &L_y^{(2)} &L_z^{(2)} \\
                             \vdots  &\vdots    &\vdots   \\
                            L_x^{(n)} &L_y^{(n)} &L_z^{(n)} 
    \end{array}\right)
    \left(\begin{array}{c}E_x \\ E_y  \\ E_z
    \end{array}\right),
\end{equation}
where the $i$-th row in the $n\times3$ matrix contains the components of antenna
effective length vector $\mathbf{L}^{(i)}$.
The matrix of antenna effective lengths, denoted $\boldsymbol{\mathsf{L}}$, has physical
dimension length and is a $n$-dimensional extension of the $\mathbf{Q}$ matrix
in \citet[eq. (2)]{Hamaker1}, for which $n=2$ and thus models dual-polarized
antennas. When $n=3$ and the antennas are linearly
independent, then they sample the full three-dimensional electric field.
Such an antenna system
is called \emph{tri-polarized} antenna in general, and tripole antenna if the
three dipoles are approximately mutually orthogonal.
One can also include antennas that sample the
magnetic field, and
an arrangement of electric and magnetic antennas can be constructed so as to
sample the full electromagnetic field at a point.
Such antennas are called electromagnetic
vector-sensors \citep{Nehorai91,Bergman05}.
In what follows, we will only be interested in
dual- or tri-polarized antenna systems.

In particular, let us consider a dual-polarized antenna that consists of two
co-located, non-mutually coupled dipole antennas,
one aligned along $\mathbf{\hat{x}}$ and the other along $\mathbf{\hat{y}}$.
The response of such a dual-polarized antenna is
\begin{equation}
  \label{eq:ideal2Dpolmeter}
  \left(\begin{array}{c} V_x \\ V_y  \end{array}
  \right)
   =L\left(\begin{array}{ccc}1 & 0 & 0 \\
                            0 & 1 & 0 
    \end{array}\right)
    \left(\begin{array}{c}E_x \\ E_y  \\ E_z
    \end{array}\right)
   =L\left(\begin{array}{c}E_x \\ E_y  \end{array}\right) ,
\end{equation}
where $L$ is the antenna effective length. We have changed the subscripts on the
voltages to reflect the right-hand side of the equation, in other words, for this
dual-polarized antenna, each voltage component is directly proportional to
a unique Cartesian component of the electric field regardless of the radiations
incidence angle.
Because of this property, it can be regarded as an ideal dual-polarized
polarimeter element.
Associated with each dual-polarized antenna element is a two-dimensional
plane in which the polarization is defined and measured.
If this plane is the same for all of the elements in a dual-polarized
interferometer or if all the planes are mutually parallel (common
design goal for polarimetric interferometers)
we will say that such an interferometer is \emph{plane-polarized},
in analogy with plane-polarized waves.
Note that a plane-polarized interferometer is not necessarily a co-planar 
interferometer, and that a non-plane-polarized interferometer may be
co-planar.
If the plane of a plane-polarized interferometer is to be specified explicitly,
we will say, e.g. in the case
of equation (\ref{eq:ideal2Dpolmeter}), that it is $xy$-polarized.

Although most existing polarimetric interferometers in radio astronomy
are intended to be plane-polarized, it is possible  to have more general antenna
elements such as electromagnetic vector-sensor arrays or electric tripole
arrays.
An real-life example of the latter is the LOIS test station,
see \citet{Bergman03,Bergman05,Bergman08,Thide04,Guthmann05}. 
The prime motivation for such a tri-polarized system is that it samples the full
electric field in a single telescope pointing rather than just a projection.

\subsection{Recovering the 2-D M.E.: dual-polarized antennas and the paraxial limit}
\label{sec:Recover2DME}

Now that we have introduced a model formalism for antenna response
we can derive the 2-D M.E. for the special but important case of the $xy$-polarized
(Hertzian dipole) interferometer, that is, the polarization plane is normal to the
centre of the FoV.   
The output voltages from the $xy$-polarized elements located at points $P_1$ and $P_2$ 
are cross-correlated and
the result can be expressed as the correlation matrix $\mathsf{R}_{ij}(1,2)=<V_{i}(1)V_{j}^\ast(2)>$
for $i,j = x,y$, where the arguments 1 and 2 refer to baseline points  $P_1$ and $P_2$.
$\boldsymbol{\mathsf{L}}$ is matrix multiplied from the left and its transpose from the right with $\briThree $,
but since it does not depend on $(l,m)$
in this case, it can be pulled out of the integral in (\ref{eq:vCZ3D}),
and so the correlator output can be written
\[
  \visTwoProj=\boldsymbol{\mathsf{L}}^{(xy)}\visThree
   \left(\boldsymbol{\mathsf{L}}^{(xy)}\right)^{\mathrm{T}} ,
\]
where
\begin{equation}
  \boldsymbol{\mathsf{L}}^{(xy)}
   =L\left(\begin{array}{ccc}
   1 & 0 & 0\\
   0 & 1 & 0\end{array}\right)
  \label{eq:dual-uv-polarized_P}
\end{equation}
is the $xy$-polarized
antenna elements, effective length matrix.
As one can see, the effect of  $\boldsymbol{\mathsf{L}}^{(xy)}$ is equivalent to projecting
the field vectors into the $xy$-plane and multiplying by $L$.
In terms of the brightness matrix the correlator output is
\begin{align}
 & \visTwoProj=|L|^2\times \nonumber \\
 & \; \mathop{\iint}_{\mathcal{F}}\boldsymbol{\mathsf{T}}^{(xy)}\briTwo
    \left(\boldsymbol{\mathsf{T}}^{(xy)}\right)^{\mathrm{T}}
   \frac{\mathrm{e}^{-\mathrm{i}2\pi\left[ul+vm
      +w\left(\sqrt{1-l^{2}-m^{2}}-1\right)\right]}}{\sqrt{1-l^{2}-m^{2}}}
   \,\mathrm{d}l\mathrm{d}m
  \label{eq:MEuv}
\end{align}
where we have introduced the $xy$-projected transformation matrix
\[
  \boldsymbol{\mathsf{T}}^{(xy)}
   = \frac{1}{L}\boldsymbol{\mathsf{L}}^{(xy)}\boldsymbol{\mathsf{T}}
   =\frac{1}{\sqrt{1-m^{2}}}\left(\begin{array}{cc}
   n & -lm\\
   0 & 1-m^2\end{array}\right)
\]
to simplify the final result,
which is now clearly two-dimensional.
Equation (\ref{eq:MEuv}) is a wide-field M.E. but with
the novel Jones matrix $\boldsymbol{\mathsf{T}}^{(xy)}$ that physically
represents a projection of the three-dimensional electric field
vector onto the $xy$-plane.

In the narrow FoV limit $\sqrt{l^2+m^2}\ll n$, so we can approximate $\boldsymbol{\mathsf{T}}^{(xy)}$ 
in equation (\ref{eq:MEuv}) up to first order in $l$ and $m$
by the two-dimensional unity matrix, and so
\begin{equation}
  \visTwoProj=|L|^2\mathop{\iint}_{\mathcal{F}}\briTwo
   \mathrm{e}^{-\mathrm{i}2\pi\left(ul+vm\right)}
   \,\mathrm{d}l\mathrm{d}m\quad\mathrm{for}\;\sqrt{l^2+m^2}\ll n .
  \label{eq:MEparaxial}
\end{equation}
This is the basic M.E. of astronomical interferometry, see \citet[eq. (14.7)]{TMS01}.
Thus we have shown that the wide-field, polarimetric vC-Z, equation (\ref{eq:vCZ3D})
indeed reduces to the usual two-dimensional, Jones vector based M.E. in the paraxial
limit.
The result, equation (\ref{eq:MEparaxial}), depended on the particular
the spherical coordinate system used as default in this paper.
Only this particular choice reduces directly to the standard 2-D M.E.
in the paraxial limit.

To see what the dual-polarized M.E. equation (\ref{eq:MEparaxial}) misses by
not including the third dimension along $z$, let us go back to
equation (\ref{eq:vCZ3D}) and let assume the paraxial approximation, $\sqrt{l^2+m^2}\ll n$.
In this case
\begin{equation}
\boldsymbol{\mathsf{T}}\approx\left(\begin{array}{cc}
1  & 0  \\
0   & 1\\
-l & -m   \end{array}\right) ,
  \label{eq:paraxialSph2cart}
\end{equation}
where we have kept only terms of first order in $l$ and $m$.
$\mathcal{V}^{(3)}_{ij} $ in equation (\ref{eq:vCZ3D})
is identical to $\mathsf{R}_{ij}/|L|^2 $ in equation (\ref{eq:MEparaxial}) for $i,j=x,y$,
but for the $z$ components,
\begin{align}
   \mathcal{V}^{(3)}_{iz}\approx-\mathop{\iint}_{\mathcal{F}}\briTwo 
   \cdot \left(\begin{array}{c} l \\ m \end{array}\right)
 \mathrm{e}^{-\mathrm{i}2\pi\left(ul+vm\right)}
   \,\mathrm{d}l\mathrm{d}m\quad\mathrm{for}\;i=x,y
  \label{eq:MEparaxialZ}
\end{align}
and
\begin{equation}
\mathcal{V}^{(3)}_{zz}\approx 0.
\end{equation}
So the $zz$ component of the three-dimensional
visibility matrix does not provide anything, but the $xz$ (and $zx$)
and the $yz$ (and $zy$) components do.
Thus even for a narrow FoV, a dual-polarized interferometer does not
measure the full set of generally non-zero electric visibilities.
Note that if the antenna array is not exactly plane-polarized,
these additional visibility components will contribute to the output-voltage
of such an array.

This leads to the important question of how serious the loss of
visibility information is in a dual-polarized interferometer.
More specifically, we ask whether
a plane-polarized interferometer, given by some $2\times 3$
matrix $\boldsymbol{\mathsf{L}}$, can in general perform
full polarimetry of an arbitrary source distribution.
The answer is that the source brightness matrix in some direction
can be determined fully only if
\[
   \det\left(\boldsymbol{\mathsf{L}\boldsymbol{\mathsf{T}}}\right)\neq 0
\]
since in this case $\boldsymbol{\mathsf{L}\boldsymbol{\mathsf{T}}}$ is invertible.
For the special but important
case $\boldsymbol{\mathsf{L}}=\boldsymbol{\mathsf{L}}^{(xy)}$
this condition is equivalent to
\[
  l^{2}+m^{2}\neq 1.
\]
Thus, an $xy$-polarized interferometer can recover the full
polarimetry except on the great circle orthogonal to $C$,
which we may call the imaging horizon.
So, in wide-field imaging with $xy$-polarized interferometers the
image horizon cannot be measured with a single telescope pointing.
Under noise-free conditions, this would pose little problem, but if we add in
the effects of noise then the great circle of directions for which
full polarimetry is not feasible broadens as a function of the
signal-to-noise ratio.

Although the discussion above was mainly focused on plane-polarized
interferometers, the three-dimensional formalism developed here can
also be applied to more general dual-polarized interferometers.
In particular it can model the situation when the polarization planes of
dual-polarized antennas in an array are not all parallel. Since several
large arrays of dual-polarized antenna based interferometers are currently
being planned for, this would address the important question of whether
such arrays should strive to be plane-polarized or whether they should purposely
not be plane-polarized to minimize the inversion problems mentioned above.

\subsection{Polarization aberration in $xy$-polarized, Hertzian dipole interferometers}

We now show that the M.E. for the $xy$-polarized, Hertzian dipole interferometer,
equation (\ref{eq:MEuv}), exhibits distortions
that depend on the look-direction, that is, the images based on these
brightnesses contain polarization aberrations \citep{McGuire90}.
This agrees with the general understanding in observational radio astronomy
that the polarimetry of a telescope is worse off-axis than on-axis.

Say we have measured the $2\times 2$ correlation matrix  $\visTwoProj (u,v,w)$.
We cannot use the formal solution (\ref{eq:vCZ3Dinv}) directly since
it is for the 3$\times$3 visibility, and it is not clear how to obtain the
remaining, unmeasured $z$ components from the $x$ and $y$ components of $\visTwoProj $.
On the other hand, for the scalar case, the formal solution for producing a
synthesized image is the following scalar, wide-field imaging equation:
\begin{equation}
 I=\frac{\sqrt{1-l^2-m^2}}{|L|^2}\iiiint \mathcal{V} e^{2\pi\mathrm{i}[ul+vm+w(n-1)]}
   \mathrm{d}u\mathrm{d}v\mathrm{d}w\mathrm{d}n ,
  \label{eq:sclWFimage}
\end{equation}
where $I=I(l,m)$ is the scalar brightness and $\mathcal{V}=\mathcal{V}(u,v,w)$
is the scalar visibility, see equation (13) in \citet{Cornwell92} who call it
the 3D method of synthesis imaging.
Let us apply (\ref{eq:sclWFimage})
to each component of  $\visTwoProj$ as if it were a scalar,
thus creating a matrix analogue of the scalar, wide-field imaging equation.
The resulting brightness matrix $\briTwoProj{xy}$ (polarized image) is
\begin{equation}
  \briTwoProj{xy}=\frac{\sqrt{1-l^2-m^2}}{|L|^2}\iiiint \visTwoProj e^{2\pi\mathrm{i}[ul+vm+w(n-1)]}
   \mathrm{d}u\mathrm{d}v\mathrm{d}w\mathrm{d}n .
  \label{eq:image2D}
\end{equation}
However, this is not the true brightness matrix because we see from
equation (\ref{eq:MEuv}) that
\begin{equation}
 \briTwoProj{xy}
   =\boldsymbol{\mathsf{T}}^{(xy)}\briTwo \left(\boldsymbol{\mathsf{T}}^{(xy)}\right)^{\mathrm{T}}
   =\frac{1}{|L|^2}\boldsymbol{\mathsf{L}}^{(xy)}\briThree \left(\boldsymbol{\mathsf{L}}^{(xy)} \right)^{\mathrm{T}} .
  \label{eq:briProjRel}
\end{equation}
It follows that $\briTwoProj{xy}$ is actually the projection of the
three-dimensional brightness matrix into
the $xy$-plane. 
The Stokes $xy$-projected brightnesses $\mathbf{S}_P$ are based directly on $\briTwoProj{xy}$
in analogy with equation (\ref{eq:StokesBright}), that is,
\begin{equation}
  \mathbf{S}_{P}(l,m)=\left(\begin{array}{c}
  I_{P}\\
  Q_{P}\\
  U_{P}\\
  V_{P}\end{array}\right)=\left(\begin{array}{c}
  B_{xx}^{(xy)}+B_{yy}^{(xy)}\\
  B_{xx}^{(xy)}-B_{yy}^{(xy)}\\
  B_{xy}^{(xy)}+B_{yx}^{(xy)}\\
  \mathrm{i}\left(B_{xy}^{(xy)}-B_{yx}^{(xy)}\right)\end{array}\right) .
  \label{eq:defStokesP}
\end{equation}
The relationship between these Stokes brightnesses and the true Stokes brightnesses $\mathbf{S}$,
which are based on $\briTwo $ can found by recasting equation (\ref{eq:briProjRel}) as
\begin{equation}
  \mathbf{S}_P=\boldsymbol{\mathsf{M}}\mathbf{S} ,
  \label{eq:StokesPdist}
\end{equation}
where
 \begin{equation}
   \boldsymbol{\mathsf{M}}=\left(\begin{array}{cccc}
\frac{1}{2}(1+n^{2}) & -\frac{m^{2}}{2}+\frac{(1+m^{2})l^{2}}{2(1-m^{2})}
    & -\frac{lmn}{1-m^{2}} & 0\\
\frac{1}{2}\left(l^{2}-m^{2}\right) 
   & 1-\frac{m^{2}}{2}-\frac{(1+m^{2})l^{2}}{2(1-m^{2})} & \frac{lmn}{1-m^{2}} 
   & 0\\
-lm & -lm & n & 0\\
0 & 0 & 0 & n\end{array}\right) .
  \label{eq:Mdist}
\end{equation}
$\boldsymbol{\mathsf{M}}$ is a Mueller matrix that quantifies the distortion of
true Stokes brightnesses based on the imaging equation (\ref{eq:image2D}).

For a very narrow FoV, $(l,m)\approx (0,0)$ and $\boldsymbol{\mathsf{M}}$ is
approximately unity. In general, however, $\boldsymbol{\mathsf{M}}$ is not the unit matrix
with the effect that the perceived Stokes vector is a distortion
of the true Stokes vector.
Thus, without further processing a plane-polarized interferometer of short dipoles will
exhibit polarization aberrations over wide fields. By contrast, a tripole array interferometer
is, at least in theory, polarimetrically aberration-free over a wide field,
since the scalar wide-field imaging equation (\ref{eq:sclWFimage})
applied to the components of its visibility matrix as if they were scalars gives $\briThree $,
which can be interpreted as the exact $\briTwo $ in Cartesian (rather than spherical) components.

\begin{figure}
\includegraphics[clip,width=1\columnwidth]{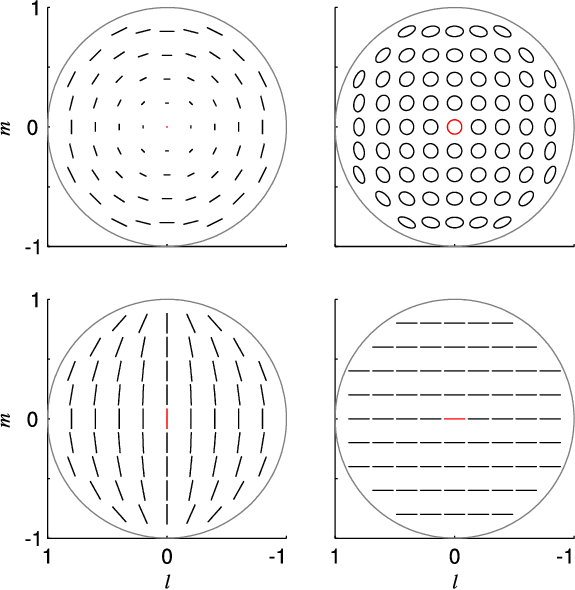}
\caption{\label{fig:PolDist4by4}Distortion of various polarized source
distributions across the hemisphere for an $xy$-polarized
interferometer. All distributions are such that the Stokes brightnesses
are constant, that is, they do not vary explicitly with direction although they may vary
implicitly due to variation of the reference system for the Stokes brightnesses,
$\boldsymbol{\hat{\theta}}$ and $\boldsymbol{\hat{\phi}}$, with direction.
The plots are of the polarization ellipses corresponding
to the normalized Stokes parameters $(Q_p/I_p,U_p/I_p,V_p/I_p)$, where
$(I_p,Q_p,U_p,V_p)$ are the $xy$-projected Stokes brightnesses as a function
of the direction $(l,m)$.
The plots clearly show polarization aberration, that is, a direction dependent distortion
in the observed polarization.
The source distributions state of polarization (shown in red) can be seen at the centre
of the FoV, $(l,m)=(0,0)$, where there is no distortion.
The upper-left panel shows $\mathbf{S}=(1,0,0,0)$,
i.e. completely unpolarized radiation, the upper-right panel shows circularly
polarized radiation, the lower-left shows radiation linearly polarized
along $\boldsymbol{\hat{\theta}}$, $\mathbf{S}=(1,1,0,0)$, 
and the lower-right panel is for radiation
linearly polarized along $\boldsymbol{\hat{\phi}}$, $\mathbf{S}=(1,-1,0,0)$ .}
\end{figure}

Examples of these wide-field distortions 
are displayed in Fig. \ref{fig:PolDist4by4}. It shows
aberration effects for source distributions
that are constant over the entire hemisphere, by which we mean
that $\briTwo (l,m)=\briTwo $, so the brightnesses do not explicitly vary with
direction\footnote{This does not mean that the distributions are necessarily
isotropic since they can still vary through their reference
to the (direction-dependent) basis $\{\boldsymbol{\hat{\phi}},\boldsymbol{\hat{\theta}}\}$.}.
Fortunately, these distortions can be compensated for in the image
plane since $\boldsymbol{\mathsf{M}}$ is invertible and well-conditioned as long
as $l^2+m^2$ is not close to one.

Another aspect of equation (\ref{eq:StokesPdist}) that is important,
is that it shows, not surprisingly, that partially polarized, wide fields
cannot be treated as scalar, wide fields.
In fact, even for the supposedly `scalar' case, that is,
when we only consider the scalar visibility $\mathcal{V}(u,v,w)=\frac{1}{2|L|^2}\mathrm{Tr}(\visTwoProj )$
of an unpolarized source distribution, so $Q=U=V=0$,
equations (\ref{eq:StokesPdist})
and (\ref{eq:image2D}) imply that
\begin{equation}
  \mathcal{V}(u,v,w)\rightleftharpoons I(l,m)\frac{2-l^2-m^2}{2\sqrt{1-l^2-m^2}} ,
  \label{eq:MyScalarvCZ}
\end{equation}
where $\rightleftharpoons$ stands for two-dimensional Fourier transform.
This should be compared
with $\mathcal{V}(u,v,w) \rightleftharpoons I(l,m)/\sqrt{1-l^2-m^2}$
for the isotropic, scalar antenna case (see, e.g., equation (1) in \citet{Cornwell92}).
This expression differs even for narrow fields since, to lowest,
non-vanishing order in $l$ and $m$, equation (\ref{eq:MyScalarvCZ}) becomes
approximately
\begin{equation}
  \mathcal{V}(u,v,w)\rightleftharpoons I(l,m)\left[ 1 -\frac{1}{4}(l^2+m^2)^2\right] ,
\end{equation}
while $\mathcal{V}\leftrightharpoons I(1-1/2(l^2+m^2))$ for the isotropic,
scalar antenna case. Thus short dipoles are less aberrated than isotropic,
scalar antennas.

The conclusion here, that scalar theory is not sufficient for the description
of the general vC-Z relations, agrees with those detailed in the field of optical
coherence, see e.g. \citet{Carter80,Saastamoinen2003}.
Also, equation (\ref{eq:MyScalarvCZ}) corresponds to an analogous equation
in \citet{Saastamoinen2003}.
Consider an unpolarized point source at $(l,m,n)=(l',m',n')$,
so $I(l,m)=I\delta(l-l',m-m')|n|$, and express $n'$ as 
$\cos\alpha$ (where $\alpha$ is the angle between the point source position
and $C$, and is equivalent to $\theta$ in \citet{Saastamoinen2003} ),
then the intensity of the unpolarized point source, as
measured by single-pixel telescope,
is $\mathcal{V}(0,0,0)=\frac{1}{2}I\left(1+\cos^2\alpha\right)$. This corresponds to
equation (47) in \citet{Saastamoinen2003}.

Although the results will
be different for other types of antennas, the Hertzian dipole is an important
special case as it is the simplest polarimetric antenna and they are directly
proportional to the Cartesian coordinates of the electric visibility matrix.

The ultimate reason for the aberration is of course that the incident,
transverse field is being projected onto the polarization plane of the
Hertzian dipoles
thereby distorting the field.
Indeed this projection is identical to the orthographic projection of a
hemisphere in cartography.
In comparison to other effects encountered in  instrumental calibration,
such as beam-shape, these effects may seem small.
They are, however, important in that they determine
the ultimate limits of polarimetry since these effects are of a
intrinsic, geometric nature.

Inspection of $\boldsymbol{\mathsf{M}}$ reveals that around $(l,m)=(0,0)$
the aberrations are $O\left(l^{2},m^{2}\right)$.
In the case of the aperture array of the SKA, the preliminary specification
\citep{Schilizzi07} has a mean FoV of 250 square degrees,
so the aberration error is in the order of \textasciitilde{}2\%,
or -16 dB, at the edge of the FoV. This should be compared with the
requirements from the key scientific projects that in some cases call for
-30 dB in polarization purity over wide fields.
Thus, these polarization aberrations will
need to be considered in wide-field imaging with SKA.

It is not difficult to show that these polarization aberrations occurring
at the edges of wide-field images also occur for off-axis imaging
with fixed mount telescopes.
This occurs, for instance, in phased arrays of crossed dipoles
such as LOFAR and the low frequency part of SKA, where imaging at scan angles away from
boresight (zenith) will be achieved by electronically steering the beam-form.
Since the crossed dipoles are fixed to the ground and are not on a mechanically rotating
mount, a situation geometrically analogous to the wide-field imaging consider above occurs.
This leads to similar aberrations in polarization for short electric dipole arrays.
Scan angles of 45$^\circ$ have been proposed for which the aberrations
will of the order of -3 dB.

\section{Conclusion}

We have derived the full set of electromagnetic vC-Z relations, which
are the basis of radio astronomical interferometry, without invoking
the paraxial approximation.
These relations allow all-sky imaging in a single telescope pointing.
We have achieved these relations by generalising the usual 2-D Cartesian Jones vector based M.E.
to a 3-D Wolf coherence matrix formulation on the (celestial) sphere.
The derived wide-field vC-Z relations are not simply trivial matrix (or vector) analogues
of the wide-field, scalar vC-Z, they exhibit direction dependent polarimetric effects.
Indeed even in the scalar limit (that is, unpolarized radiation), our M.E. is not the same as
the M.E. derived from scalar theory, in the case of Hertzian dipoles.
Furthermore, we found that, for an arbitrary wide-field of sources, the electric
visibilities generally have nine complex components for an arbitrary baseline.
This implies that the standard Stokes (Cartesian) visibilities do not
provide a full description of electric coherence for wide fields.
We have also shown that our vC-Z relation (for the electric field) reduces to
the standard 2-D M.E. after a 2-D projection.
We showed that a consequence of this projection is that plane-polarized,
Hertzian dipole interferometers are aberrated polarimetrically.
Fortunately, these aberrations can be corrected for
in the image plane for sources sufficiently far away from the
plane of the dual-polarized antenna elements.

Besides its use in the derivation of the wide-field vC-Z relation,
we believe that the 3-D Wolf formalism can be useful in constructing
more general 3-D M.E. in cases requiring the full set of electric components.
As examples, we mention the modelling of tri-polarized element arrays,
the modelling of dual-polarized element arrays that are not strictly plane-polarized
(due to manufacturing errors or the Earth's curvature), and the modelling of
propagation that is not along the line-of-sight (due to refraction or diffraction
in the ionosphere, e.g.).

\section*{Acknowledgements}

We thank the referee, J. P. Hamaker, for valuable and constructive comments.
This work is supported by the European Community Framework Programme
6, Square Kilometre Array Design Studies (SKADS), contract no 011938,
and the Science and Technology Facilities Council (STFC), UK.

\bibliographystyle{mn2e}

\appendix

\section{Full electromagnetic \vCZ\ relations}
\label{EMvCZ}

In the previous sections we considered only the electric field
and its auto-correlation, but now we look at the full electromagnetic field.
One motivation to do this is that, especially
for low radio frequencies, it is possible to sample both the electric and
the magnetic fields and thereby measure electromagnetic coherence fully,
see \citet{Bergman03}, \citet{Bergman05} or \citet{Bergman08}.
Such sensors have been deployed in some radio interferometers
and will possibly provide unique and novel astronomical measurements. 
For completeness then, we present the
rest of the electromagnetic correlations analogous to the electric vC-Z relation
in equation (\ref{eq:vCZ3Drel}).

We found previously that the $3\times 3$ electric visibility matrix was related
to the $2\times 2$ electric brightness matrix as 
\begin{equation}
  \boldsymbol{\mathcal{V}}^{(EE)}(u,v,w)\leftrightarrows\briTwo^{(ee)}(l,m)
  \label{eq:vCZsymbolicE}
\end{equation}
where $\leftrightarrows$ denotes the vC-Z relationship given
by equation (\ref{eq:vCZ3D}).
Note that we have now changed the notation of
$\visThree$ to $\boldsymbol{\mathcal{V}}^{(EE)}$ and $\briTwo$
to $\briTwo^{(ee)}$ to indicate that these quantities represent auto-correlations
of visibility electric field and the brightness Jones vectors, respectively.
A full electromagnetic vC-Z relationship between brightnesses
and visibilities requires also the auto-correlations of
the magnetic field and the cross-correlation between
the electric and magnetic fields,
see \citet[chap. 6]{Mandel95}. We define the electromagnetic visibilities for
baseline $\mathbf{D}_{\lambda}$ with respect to the phase reference direction $\mathbf{s}_0$ as
\begin{align}
  \mathcal{V}^{(EE)}_{ij}(\mathbf{D}_{\lambda})&=\left\langle E_i(\mathbf{r})E^\ast_j(\mathbf{r}-\mathbf{D})\right\rangle
   \exp(\mathrm{i}2\pi \mathbf{s}_0 \cdot\mathbf{D}_{\lambda}) \\
  \mathcal{V}^{(EH)}_{ij}(\mathbf{D}_{\lambda})&=\left\langle E_i(\mathbf{r})H^\ast_j(\mathbf{r}-\mathbf{D})\right\rangle
 \exp(\mathrm{i}2\pi \mathbf{s}_0 \cdot\mathbf{D}_{\lambda})  \\
  \mathcal{V}^{(HE)}_{ij}(\mathbf{D}_{\lambda})&=\left\langle H_i(\mathbf{r})E^\ast_j(\mathbf{r}-\mathbf{D})\right\rangle
 \exp(\mathrm{i}2\pi \mathbf{s}_0 \cdot\mathbf{D}_{\lambda}) \\
  \mathcal{V}^{(HH)}_{ij}(\mathbf{D}_{\lambda})&=\left\langle H_i(\mathbf{r})H^\ast_j(\mathbf{r}-\mathbf{D})\right\rangle 
 \exp(\mathrm{i}2\pi \mathbf{s}_0 \cdot\mathbf{D}_{\lambda}) 
\end{align}
for $i,j=x,y,z$, where $\mathbf{H}(\mathbf{r})$ is the magnetic field at $\mathbf{r}$.
The electromagnetic brightnesses in direction
$\mathbf{s}$ are defined as
\begin{align}
  B^{(ee)}_{ij}(\mathbf{s})&=\left\langle e_i(\mathbf{s})e_j^\ast(\mathbf{s}) \right\rangle \\
  B^{(eh)}_{ij}(\mathbf{s})&=\left\langle e_i(\mathbf{s})h_j^\ast(\mathbf{s}) \right\rangle \\
  B^{(he)}_{ij}(\mathbf{s})&=\left\langle h_i(\mathbf{s})e_j^\ast(\mathbf{s}) \right\rangle \\
  B^{(hh)}_{ij}(\mathbf{s})&=\left\langle h_i(\mathbf{s})h_j^\ast(\mathbf{s}) \right\rangle ,
\end{align}
for $i,j=\phi ,\theta$, where $\mathbf{h}(\mathbf{s})$ is the magnetic field from the source distribution in
direction $\mathbf{s}$.

The magnetic field associated with the visibilities can be derived by applying
Faraday's law,
\[
\mathbf{H}(\mathbf{r},t)
   =-\mathrm{i}\frac{c}{2\pi\nu Z_0}\boldsymbol{\nabla}\boldsymbol{\times}
    \mathbf{E}(\mathbf{r},t) ,
\]
where $Z_{0}$ is the impedance of free space, to the electric field used in derivation
in section \ref{sec:WolfvCZ}. The result is that the magnetic analogues of the 
electric vC-Z expressions involve a magnetic Jones
vector $\mathbf{h}=( h_\phi , h_\theta )^{\mathrm{T}}$ that is related to the electric Jones vector through
\[
  -\frac{1}{Z_{0}}\mathbf{s}\times
  \left( \boldsymbol{\mathsf{T}}\mathbf{e}\right)
  =\boldsymbol{\mathsf{T}}\boldsymbol{\mathsf{F}} \mathbf{e}
  =\boldsymbol{\mathsf{T}}\mathbf{h}
\]
where
\begin{equation}
   \boldsymbol{\mathsf{F}}=\frac{1}{Z_0}\left(\begin{array}{cc} 0 & 1 \\ -1 & 0 \end{array}\right) ,
\end{equation}
or in other words
\[
 \mathbf{h}
 =\frac{1}{Z_{0}}\left(\begin{array}{c} +e_\theta  \\ -e_\phi \end{array}\right) .
\]
This says that the magnetic
Jones vector is directly determined from the electric Jones vector, and so no
other independent electromagnetic source coherence statistics exist (in the far-field
zone) other than the $2\times 2$ electric brightness matrix.

Repeating the derivation in section \ref{sec:WolfvCZ} but for the magnetic field
we find we can write the rest of the vC-Z relations as
\begin{align}
  \boldsymbol{\mathcal{V}}^{(EH)}
   &\leftrightarrows \briTwo^{(eh)} 
   =\briTwo^{(ee)}\boldsymbol{\mathsf{F}}^{\mathrm{T}} \\
  \boldsymbol{\mathcal{V}}^{(HE)}
   &\leftrightarrows \briTwo^{(he)}
   =\boldsymbol{\mathsf{F}} \briTwo^{(ee)} \\
  \boldsymbol{\mathcal{V}}^{(HH)}
   &\leftrightarrows \briTwo^{(hh)}
   =\boldsymbol{\mathsf{F}} \briTwo^{(ee)}\boldsymbol{\mathsf{F}}^{\mathrm{T}} .
\end{align}
Thus one can see $\boldsymbol{\mathsf{F}}$ as a Jones-like matrix that switches
between electric and magnetic coherencies.

These relations provide a complete description of the second-order
coherence of the electromagnetic radiation. So, for instance, one
can easily compute the Poynting visibility vector using the above relations,
\begin{align}
   &\left\langle \Re\left[\mathbf{E}\left(\mathbf{r}\right)
     \times\mathbf{H}^{\ast}\left(\mathbf{r}-\mathbf{D}\right)\right]\right\rangle 
  = \nonumber \\   &\quad
-\frac{1}{Z_{0}}\iint I\mathbf{s} 
  \mathrm{e}^{-\mathrm{i}2\pi\left[ul+vm+w\left(n
     -1\right) \right]} \, \mathrm{d}\Omega
\end{align}
where $I=I(l,m)=|e_\theta|^2+|e_\phi|^2$.
This says that the power flux visibility vector
is vC-Z related to the Stokes brightness propagating from sources.

\section{Coordinate transformations}
\label{sec:CoordTrans}

In this paper we have used the standard definitions
of the $uvw$- and $lmn$-spaces, see \citet{TMS01}, and we have used
a spherical basis $\{\boldsymbol{\hat{\theta}},\boldsymbol{\hat{\phi}}\}$.
Both these coordinate systems are relative to the centre of the FoV.
Often one wants to transform
to some other system, and some details can be found in the standard texts \citep{TMS01}.
However, what is not usually done explicitly is the transformation
of the non-scalar brightnesses and visibilities.
We will now show how to rotate the matrix brightnesses and visibilities
found in the polarimetric, wide-field vC-Z, equation (\ref{eq:vCZ3D}).

Equation (\ref{eq:vCZ3D}) uses the Cartesian $\{\mathbf{\hat{x}},\mathbf{\hat{y}},\mathbf{\hat{z}}\}$ system
as a basis for $\visThree $, $\mathbf{s}$, $\mathbf{D}_{\lambda}$
and the rows of $\boldsymbol{\mathsf{T}}$. It uses the spherical 
base vectors $\{\boldsymbol{\hat{\theta}},\boldsymbol{\hat{\phi}}\}$
as a basis for $\briTwo $ and the columns of $\boldsymbol{\mathsf{T}}$.
Under a rotation given by the
3$\times$3 orthogonal matrix $\boldsymbol{\mathsf{Q}}$, the vectors and matrices
in the Cartesian system transform in the usual manner:
\begin{align}
  \mathbf{s}'&=\boldsymbol{\mathsf{Q}} \mathbf{s} , \label{eq:rots} \\
  \mathbf{D}_{\lambda}'&=\boldsymbol{\mathsf{Q}}\mathbf{D}_{\lambda} , \\
  \boldsymbol{\mathcal{V}}'^{(3)}(\mathbf{s}')&=\boldsymbol{\mathsf{Q}}
            \visThree(\mathbf{s}')
    \boldsymbol{\mathsf{Q}}^{\mathrm{T}} .
  \label{eq:vis3trans}
\end{align} 
Note that a transformation analogous to equation (\ref{eq:vis3trans})
for the $2\times 2$ matrix $\mathbf{\mathsf{R}}$, discussed in section \ref{sec:Recover2DME}, does not exist
since $\mathbf{\mathsf{R}}$ lacks one of the dimensions necessary for a general coordinate
transformation.
Only the correlator output from a tri-polarized antenna array can be
fully transformed in general.

The brightness matrix rotates as
\begin{equation}
  \briTwo '(\mathbf{s}') =\boldsymbol{\mathsf{T}}^{\mathrm{T}}(\mathbf{s}')
                \boldsymbol{\mathsf{Q}}
             \boldsymbol{\mathsf{T}}(\mathbf{s})
               \briTwo(\mathbf{s}) 
                   \boldsymbol{\mathsf{T}}^{\mathrm{T}}(\mathbf{s})
           \boldsymbol{\mathsf{Q}}^{\mathrm{T}}\boldsymbol{\mathsf{T}}(\mathbf{s}'),
  \label{eq:bri2trans}
\end{equation}
where $\mathbf{s}$ on the left-hand side should be converted into $\mathbf{s}'$
using equation (\ref{eq:rots}).

Actually, $\boldsymbol{\mathsf{Q}}$ need not be the same in
equations (\ref{eq:vis3trans}) and (\ref{eq:bri2trans}),
since the spherical and Cartesian systems can be rotated separately.
This can be used to change the relative alignment (rotation) of the spherical
system relative the Cartesian system. The net result is a change in matrix $\boldsymbol{\mathsf{T}}$
that relates the spherical components to the Cartesian components.
Consider the special case when the spherical system is rotated in the positive sense
around the $\mathbf{\hat{x}}$-axis through angle $\Theta$ relative the Cartesian system.
The effect on the vC-Z relations is that $\boldsymbol{\mathsf{T}}$
is replaced by $\boldsymbol{\mathsf{T}}'$, where
\begin{align}
  \boldsymbol{\mathsf{T}}'=&\frac{1}{\sqrt{1-(m\cos\Theta-n\sin\Theta)^2}} \\
    &\times\left(\begin{array}{cc}
     n\cos\Theta+m\sin\Theta   &-lm\cos\Theta +ln\sin\Theta  \\
     -l\sin\Theta               & (1-m^2)\cos\Theta +mn\sin\Theta\\
     -l\cos\Theta               & -mn\cos\Theta+(n^2-1)\sin\Theta\end{array}\right) .
  \label{eq:sph2cart_gen}
\end{align}
Obviously, for $\Theta=0$ we
obtain $\boldsymbol{\mathsf{T}}'=\boldsymbol{\mathsf{T}}$, which is the matrix used in most of this paper.

At the field centre, $l=m=0$, so
\begin{equation}
   \boldsymbol{\mathsf{T}}'=\left(\begin{array}{cc} 
     1 & 0 \\
     0 & 1 \\
     0 & 0 \end{array}\right)
\end{equation}
for all $\Theta\neq \pi /2$.
However, the $x$ and $y$ components of its first derivatives
are zero only for $\Theta=0$, that is 
\[
  \left. \frac{\partial\mathsf{T}'_{ij}}{
    \partial l }\right|_{l=m=0}=0,
  \quad
 \left. \frac{\partial\mathsf{T}'_{ij}}{
    \partial m }\right|_{l=m=0}=0,
   \;\mathrm{for}\begin{cases} i=x,y & \\ j=\theta,\phi & \end{cases}
\]
only for $\Theta=0$. Thus, the special case we have used in this
paper,  $\boldsymbol{\mathsf{T}}'=\boldsymbol{\mathsf{T}}$, can be said
to possess a projection ($xy$) that is locally flat at the field centre.
This is a reason for choosing the spherical system with $\Theta=0$ as a default,
since all other cases would lead to a 2-D M.E. (\ref{eq:MEparaxial}) with additional first-order terms
that account for the coordinate system curvature within the FoV.

The matrix $\boldsymbol{\mathsf{T}}'$ can be used to adapt the vC-Z relations given
in this paper, such as equation (\ref{eq:vCZ3D}), to standard celestial coordinate systems
such as the equatorial system or the azimuth-elevation system.
In essence, assuming that the pole $A$ is towards
the Earth's North pole (in the case of equatorial coordinates) or zenith (in the case of
az-el coordinates), one simply performs
$ \boldsymbol{\mathsf{T}} \rightarrow \boldsymbol{\mathsf{T}}'$ and then interprets
$\Theta $ as either the declination (equatorial case) or the elevation (az-el case) of the centre of the FoV.
Note however that the Cartesian coordinates, in which the $uvw$ and $lmn$ spaces are
expressed, would still need to transformed for complete agreement with these standard celestial
systems, but for this common task we refer to standard texts such as \citet{TMS01}.

\label{lastpage}
\end{document}